\documentclass[12pt]{article}
\usepackage{amsmath,
graphicx, enumerate}
\usepackage[letterpaper,top=0.5in, bottom=1in, left=0.8in, right=0.8in, headsep=0.in, centering]{geometry}
\usepackage{amssymb}
\usepackage{epsfig}
\title{Wave-Particle Duality and  the Hamilton Action
}

\author{Gregory I. Sivashinsky\\  
Sackler Faculty of Exact Sciences,
School of Mathematical Sciences, \\
Tel Aviv University, Tel Aviv 69978, Israel}
\date{}
\begin{document}
\maketitle
\begin{center}
{\bf Abstract} 
\end{center}
\noindent The Hamilton-Jacobi equation of relativistic quantum mechanics is revisited.  The equation 
is shown to permit
 solutions in the form of breathers (oscillating/spinning solitons), displaying simultaneous particle-like and wave-like behaviour.

\bigskip


\medskip


\setcounter{equation}{0}
%\section{Introduction}

\bigskip

\bigskip

\noindent A mathematical representation of the dual wave-particle nature of light and matter remains one of the major challenges of quantum theory (e.g.[1-6]).  The present letter is an attempt to resolve this issue through an appropriately revised  Hamilton-Jacobi formalism.

\noindent Consider the classical Klein-Gordon (KG) equation for a free particle,
\begin{eqnarray} \label{1}
\Box \Psi+\left( mc/\hbar \right)^2  \Psi=0, \qquad \left(\Box=(1/c^2)\partial^2/\partial t^2-\nabla^2 \right)
\end{eqnarray}

%\noindent Equation (\ref{1}) is obtained from the original KG equation by merely doubling the
%Plank constant  $\hbar$.               The major impact of this seemingly minor change will become clear shortly.

\noindent Upon the familiar substitution,
\begin{eqnarray}\label{2}
\Psi=\exp \left( iS/ \hbar \right)
\end{eqnarray}
the         KG equation (1) converts into the    quantum Hamilton-Jacobi (QHJ) equation for the Hamilton's action-function     $S$,
\begin{eqnarray}\label{3}
\left( 1/c^2\right)\left(\partial S/ \partial t \right)^2-\left( \nabla S\right)^2=i \hbar \Box S+m^2 c^2
\end{eqnarray}
The        KG equation allows for a two-term spherically-symmetric solution
\begin{eqnarray}\label{4}
\Psi=\exp\left[  -i\left(\frac{mc^2}{\hbar}\right) t\right]
+\alpha \exp\left[-2i\left(\frac{mc^2}{\hbar} \right)t\right] j_0\left[ \sqrt{3}\left(\frac{mc}{\hbar}\right)r\right],
\end{eqnarray}
where       $r=\sqrt{x^2+y^2+z^2}$,~ $\alpha$  is a parameter assumed to be prescribed, and
\begin{eqnarray}
j_0=\frac{\sin\left( \sqrt{3}\left({mc}/{\hbar}\right)r\right)}{\sqrt{3}\left({mc}/{\hbar}\right)r}
\end{eqnarray}
is the zero-order spherical Bessel function.

\noindent The second term in Eq. (\ref{4}) is a standing spherically-symmetric breather,  a localized periodically oscillating structure.
In terms of the action-function $S$, by virtue of (\ref{2}), Eq. (\ref{4}) readily yields,
\begin{eqnarray}\label{6}
S=-mc^2t-i\hbar \ln\left\{1+\alpha\exp\left[-i\left( \frac{mc^2}{\hbar}\right)t\right] j_0\right\}
\end{eqnarray}
Here the first term corresponds to the classical action-function for a free particle in the rest-system, while the second term represents its localized excitation, oscillating and {\it non-spreading}.   The frequency of the oscillations is set precisely at    $mc^2/\hbar$, in line with the de Broglie theory.  Note that here the frequency is not affected by the nonlinearity of the system, preserving its value irrespective of the breather intensity.
\noindent Away from the breather's core ($r>>\hbar/mc$ ),
\begin{eqnarray}\label{7}
S=-mc^2t-i\alpha\hbar\exp\left[-i\left(\frac{mc^2}{\hbar}\right)t\right] j_0
\end{eqnarray}
The oscillations are therefore asymptotically {\it monochromatic}, again in accord with the de Broglie picture.
%\noindent Represent Eq. (\ref{6}) as
%\begin{eqnarray}\label{8}
%S={\rm Re} (S)+i{\rm Im} (S)
%\end{eqnarray}
%As may be readily shown,
%\begin{eqnarray}\label{9}
%{\rm Re} (S)=-mc^2 t-2 \hbar {\rm arctg} \left[ \frac{\alpha j_0\sin(mc^2t/\hbar)}{1+\alpha j_0\cos(mc^2t/\hbar)}\right]
%\end{eqnarray}
%and
%\begin{eqnarray}\label{10}
%{\rm Im}  (S)=-\hbar\ln\left [\left(1+\alpha j_0\cos (mc^2 t/\hbar)\right)^2+\alpha^2j_0^2\sin^2(mc^2t/\hbar)\right]
%\end{eqnarray}
%So far the parameter   $\alpha$         remains undetermined.  There are therefore infinitely many suitable breathers, which is typical of a non-dissipative system such as (\ref{3}).  We observe, however, that while      ${\rm Re} (S)$        is bounded for any       $\alpha$, ~ ${\rm Im} (S)$       acquires  logarithmic singularities at             $|\alpha| \ge 1 $             when  $\cos\left(mc^2 t/\hbar\right)=\pm 1$.
%This allows one to anticipate that the associated marginal solutions ($|\alpha|=1$) are in some sense 'privileged'. In any case, 
Similar to oscillations of an ideal pendulum,  the breather (\ref{6}) is stable to small perturbations.  
The stability here is understood in a weak (non-asymptotic) sense.
%The issue definitely calls for further analysis.

\noindent Until now we have dealt with a particle at rest.  For a particle moving at a constant velocity         $v$     along, say,        $x$  - axis, the corresponding expression for the action-function is readily obtained from Eq. (\ref{6}) through the Lorentz transformation,
\begin{eqnarray} \label{11}
t\to \frac{t-xv/c^2}{\sqrt{1-(v/c)^2}}, \qquad x\to\frac{ x-vt}{\sqrt{1-(v/c)^2}}.
\end{eqnarray}

\noindent The transformed Bessel function      $j_0$          will then mimic the motion of the classical particle while the transformed temporal factor $\exp\left[-i (mc^2/\hbar)t\right]$ will turn into the associated 
de Broglie wave, thereby ensuring simultaneous particle-like and wave-like behaviour.   Moreover, unlike conventional quantum mechanics, here the de Broglie wave acquires the clear {\it deterministic }meaning of being simply an excitation of the action-function     $S$, a complex-valued potential in configuration space.

\noindent The outlined formalism seems fully compatible with the Ôdouble-slitÕ experiment.  The particle 'feels' the distant environment through its vibrating 'tail' (\ref{7}), which navigates the particle in compliance with the far-field geometry. Note, that while the particle velocity is clearly subluminal
 its communication with the environment is {\it superluminal}, which is not incompatible with the Lorentz-invariance  of the system.

\noindent If, as is conventional, we associate the gradients    $-\partial S/\partial t, \nabla S$                            
with the particle energy         $E$        and momentum      ${\bf p}$, then the Einstein relation
$E=c\sqrt{p^2+m^2c^2}$
appears to hold only   far from the     $\hbar/mc$ - wide breather's core.  
The correspondence with classical relativistic mechanics  is therefore complied with. Note, that for a particle at rest the relation $E=mc^2$, apart from $r>>\hbar/mc$, holds  also {\it on average}
over the entire breather.

\noindent Equation (\ref{3}) is also applicable to massless particles (photons, neutrinos).  For $m\to 0$,  $v\to c$, 
and finite       $E$, the breather structure of the associated action-function is fully preserved.

\noindent For a particle moving in a field $(U,{\bf A})$, the terms  $\partial S/\partial t$, $\nabla S$
 in Eq. (\ref{3}) should be changed to       $\partial S/\partial t+U$, $\nabla S-{\bf A}/c$.                               
Assuming that $(U,{\bf A})$ satisfies the familiar Lorentz calibration condition,
\begin{eqnarray}\label{12}
\left(1/c\right)\partial U/\partial t+{\rm div}~{\bf A}=0
\end{eqnarray}
the field-version of the            QHJ equation takes on the following compact form,
\begin{eqnarray}\label{13}
\left( 1/c^2\right)\left( \partial S/\partial t+U\right)^2-\left(\nabla S- {\bf A}/c \right)^2=
m^2c^2+i\hbar\Box S
\end{eqnarray}

\noindent In addition to spherically symmetric breathers, Eqs. (\ref{1}) (\ref{3}) also permit  asymmetric breathers, spinning about  some axis.  In the latter case the second term of Eq. (\ref{4}) should be replaced by
\begin{eqnarray}\label{14}
\alpha\exp\left[-2i\left(\frac{mc^2}{\hbar}\right)t+in\phi\right]j_l\left[\sqrt{3}\left(\frac{mc}{\hbar}\right)r\right]P_l^n\left(\cos\theta\right),
\end{eqnarray}							
where       $j_l$, $P_l^n$    are high-order spherical Bessel functions and associated Legendre polynomials (notations are conventional).  It would be interesting to ascertain in what way (if any) the double-valued spinning breather, emerging at $n=l=1/2$,  may be linked to the Dirac wave-function.

\noindent The next question is how to reproduce quantization directly in terms of the breathing  action-function.  The geometrically simplest situation, where such an effect might manifest itself, is the periodic motion of an otherwise free particle over a closed interval   $0<x<d$.  In this case Eq. (\ref{3}) must be considered jointly with two boundary conditions,
\begin{eqnarray}\label{15}
&& \partial S(0,y,z,t)/\partial t=\partial S(d,y,z,t)/\partial t, \nonumber \\
&& \partial S(0,y,z,t)/\partial x=\partial S(d,y,z,t)/\partial x
\end{eqnarray}
Any classical action-function for a free particle,
\begin{eqnarray}\label{16}
 S=-Et+px  
%&& \left( E=c\sqrt{p^2+m^2c^2}, p=mv/\sqrt{1+(v/c)^2} \right)
\end{eqnarray}
is clearly a solution of this problem.  However, in the case of a locally excited action the situation may prove different.  Thanks to the boundary conditions (\ref{15}), the moving breather will interact with itself, and this may well lead to its self-destruction unless        some particular conditions are met.                                         
%Thus, unlike the classical solution (\ref{16}), a locally excited action-field may not be adjustable to all external conditions.   Further exploration of this intriguing question is needed.

\noindent Consider first the simplest case of a particle at rest $(v=0)$.  The pertinent solution is readily obtained by converting the problem for a finite interval (\ref{3}) (\ref{15}) into a problem for an infinite interval ($-\infty<x<\infty$) filled with a      $d$-periodic train of standing breathers of the type described previously.  The resulting action-function reads,
\begin{eqnarray}\label{17}
S=-mc^2t-i\hbar\ln\left\{ 1+\alpha\exp\left[-i\left(\frac{mc^2}{\hbar}\right)t\right]\sum_{k}\left( j_0\right)_{d,0}^{(k)}\right\},
\end{eqnarray}
where 										
\begin{eqnarray}\label{18}
\left(j_0\right)_{d,0}^{(k)}=\frac{\sin\left(\sqrt{3}\left(mc/\hbar\right)r_{d,0}^{(k)}\right)}{\sqrt{3}\left(mc/\hbar\right)r_{d,0}^{(k)}},
\end{eqnarray}
\begin{eqnarray}\label{19}
r_{d,0}^{(k)}=\sqrt{(x-kd)^2+y^2+z^2} \qquad (k=0,\pm 1,\pm 2,...)
\end{eqnarray}
Here the second subscript stands for    $v=0$.     

\noindent The action-function for a moving particle $(v\ne0)$  is obtained from (\ref{17}) (\ref{18}) (\ref{19}) through the Lorentz transformation (\ref{11}), provided  $d$  is replaced by   $d/\sqrt{1-(v/c)^2}$. The latter step is needed to balance the relativistic contraction, and thereby to preserve the spatial period ($d$)  of the system.  The resulting action-function thus reads,
\begin{eqnarray}\label{20}
  S=-Et+px-i\hbar\ln \left\{1+\alpha\exp\left[i\left(\frac{-Et+px}{\hbar}\right)\right]\sum_{k}\left(j_0\right)_{d,v}^{(k)}\right\},         
\end{eqnarray}
where
\begin{eqnarray}\label{21}
\left(j_0\right)_{d,v}^{(k)}=\frac{\sin\left(\sqrt{3}\left(mc/\hbar\right)r_{d,v}^{(k)}\right)}{\sqrt{3}\left(mc/\hbar\right)r_{d,v}^{(k)}},
\end{eqnarray}
\begin{eqnarray}\label{22}
r_{d,v}^{(k)}=\sqrt{\left(\frac{x-vt-kd}{\sqrt{1-(v/c)^2}}\right)^2+y^2+z^2}.
\end{eqnarray}
The spatial 	$2\pi\hbar/p$ - periodicity of  $\exp\left[i\left(-Et+px\right)/\hbar\right]$								
is compatible with the spatial     $d$   - periodicity of      
%$\sum_k\left(j_0\right)_{d,v}^{(k)}$                           
$\sum_{k}\left(j_0\right)_{d,v}^{(k)}$
 only if
\begin{eqnarray}
dp=2\pi n\hbar \qquad (n=1,2,3,...),
\end{eqnarray}
which recovers the familiar Bohr-Sommerfeld quantum  condition.

\noindent At this stage it is difficult to see whether the amended QHJ formalism is indeed rich enough to reproduce all the basic features of quantum-mechanical phenomenology.  In any case, a few preliminary observations already show that a mathematical representation of unified wave-particle behaviour 
through the conventional QHJ equation 
is not unfeasible.

\medskip

\noindent {\bf Acknowledgements}

\noindent Helpful discussions with Irena Brailovsky are gratefully acknowledged .

\noindent These studies were supported in part by the Bauer-Neumann Chair in Applied Mathematics and Theoretical Mechanics, the US-Israel Binational Science Foundation (Grant 2006-151), and the Israel Science Foundation (Grant 32/09).

\end{document}